\documentclass[12pt]{article}

\usepackage{graphics} 
\usepackage{graphicx}
\usepackage{epsfig}

\def\b{\beta}

\def\G{\Gamma}
\def\d{\delta}
\newcommand{\ov}{\overline v}

\newcommand{\be}{\begin{equation}}
\newcommand{\ee}{\end{equation}}
\newcommand{\bea}{\begin{eqnarray}}
\newcommand{\eea}{\end{eqnarray}}

\begin{document}

\hskip 12cm NSF-KITP-12-009

\begin{center}
\center{{\Large Lattice Gauge Theory - Gravity duality and Coulomb's constant in five dimensions}
\vskip 3cm
{\large Nikos Irges$^{1,2}$}
\vskip .5cm
1. {\it Department of Physics, National Technical University of Athens\\ 
         GR-15780 Athens, Greece}

2. {\it KITP Santa Barbara, Santa Barbara CA 93106, U.S.A.}}

\vskip 2cm

{\bf Abstract}
\end{center}

The purpose of this paper is to perform a quantitative check of
gauge theory - gravity duality in a nonconformal, nonsupersymmetric context.
In order to do so we define $k_5$, an object extracted from the Wilson Loop, that plays the role of 
Coulomb's constant for $SU(N)$ gauge theories in five dimensions
and we argue that one of its virtues is that it could be minimally sensitive to $N$.
This allows us to compute $k_5$ on one hand 
from the gravitational backreation of a large number $N$ of $D4$-branes,
and on the other from a lattice mean-field expansion for $N=2$.
We find a $2\%$ numerical agreement between the two approaches.

\newpage

\section{Introduction}

Most of the evidence for gauge - gravity duality comes from conformal systems with some amount of
supersymmetry. It is a challenge to come up with examples where the duality is at work
while these symmetries are absent or broken. Another obstruction in collecting evidence for the duality is that it is best understood when 
the side of the gauge theory (we restrict ourselves here to $SU(N)$ gauge theories) is strongly coupled
and $N$ is large. The strong coupling requires nonperturbative methods among which our analytical tools are 
restricted in four or higher dimensions. 
Numerical methods on the other hand become prohibitive for these systems when $N$ is really large.
In this paper we present an example in five space-time dimensions, where all the above difficulties may be possible to overcome. 
In order to construct the example we first define a certain object that we name Coulomb's constant,
an appropriate generalization of our familiar constant in four dimensions.

Coulomb's law in four dimensions for a unit positive charge is typically written in the form $E=1/4\pi (1/r^2)$
(we work in units where $\epsilon_0=1$). The $1/4\pi$ is a convention suggested by Gauss's law and it represents,
in these units, the classical overall strength of the underlying $U(1)$ force.
One refers to this constant as Coulomb's constant.
A generalization of this definition to any quantum $SU(N)$ theory requires the knowledge of 
three dimensionless objects: the charge $c_1(r)$
obtained from the static potential $V_4(r)$, the 't Hooft coupling $\lambda = g^2 N$ 
with $g$ the renormalized gauge coupling and the 
group-dependent Casimir index $C_F=(N^2-1)/2N$. Then, the combination
\be
k_4=\frac{N}{C_F}\frac{c_1}{\lambda} = \frac{1}{4\pi}\label{4dC}
\ee
is a nonperturbative and $N$-independent definition of Coulomb's constant\footnote{Here and in the 
rest of this letter the term "Coulomb's constant" should be taken with a grain of salt. 
What this really means is that one can define a renormalized coupling 
through $c_1(r)$ anywhere on the phase diagram, such that it satisfies eq. (\ref{4dC}).} in four dimensions.
Renormalizability protects the validity of this definition throughout the phase diagram.
Notice in particular that the quantity defined in Eq. (\ref{4dC}) is scheme-independent in perturbation theory.
To illustrate the point, we write it in two different schemes, the '$c$' and '$qq$' schemes  \cite{FrancRainer}:
\bea
\frac{1}{4\pi} &=& \frac{c_1^{(c)}(r)}{g_c^2(1/r)C_F} = \left[\frac{c_1(r)}{g^2(1/r)}\right]_{\rm c-scheme}\frac{1}{C_F}\nonumber\\
\frac{1}{4\pi} &=& \frac{c_1^{(qq)}(r)}{g_{qq}^2(1/r)C_F} = \left[\frac{c_1(r)}{g^2(1/r)}\right]_{\rm qq-scheme}\frac{1}{C_F}
\eea
where $c_1^{(c)}(r)=-1/2 r^3 V_4''(r)$, $c_1^{(qq)}(r)=r^2 V_4'(r)$. 
We denote $V'_4(r)=\partial V_4(r)/\partial r$ etc.
The couplings in the two different schemes are related in a well defined way in perturbation theory,
as is their relation to more conventional in the continuum schemes such as the ${\rm \overline {MS}}$ scheme \cite{FrancRainer}.
Clearly, while the charges $g^2$ and the $c_1$'s are scheme dependent, their ratio, i.e. our quantity of interest, is not.

In five dimensions things are more complicated as the classical $U(1)$ Coulomb constant $1/(2\pi^2)$, suggested by Gauss's law, proves
to be more difficult to generalize. To begin, $SU(N)$ gauge theories are nonrenormalizable in five dimensions.
As a consequence, in the non-self-interacting or perturbative limit they are free and in its vicinity cut-off dominated.
To define nonperturbatively a computationally tractable interacting theory one observes that
even in infinite volume these theories possess a first order phase transition separating a confined phase at strong coupling
from a Coulomb phase at weak coupling and that as the phase transition is approached
(from the side of the Coulomb phase in this work), cut-off effects diminish \cite{FM}. 
By analogy to eq. (\ref{4dC}) we form the dimensionless combination
\be
k_5=2 \frac{c_2}{\lambda_5}\label{5dC}
\ee
where $c_2$ is the static charge of dimension length appearing in the static potential  $V_5(r)={\rm const.} - c_2/r^2$
and $\lambda_5=g_5^2 N$ the five-dimensional analogue of $\lambda$ which has also dimension of length due to the 
dimensionful five-dimensional coupling $g_5$. 
As for the dimensionless constant, instead of $N/C_F$ used in four dimensions, we take its infinite $N$ limit, 2.
In this letter, we take $k_5$ as our definition of Coulomb's constant in five dimensions and 
we compute it in two different ways.
We stress that in five dimensions there is no unambiguous notion of perturbative "schemes"
because the theory is nonrenormalizable. 
Nevertheless, our quantity is expected to be independent of any possible scheme definition, 
for the same reasons as in four dimensions.
Moreover, since our computations are anyway far from the perturbative regime
and in addition we compute exactly the same quantity with both methods, we expect our comparative study
of Coulomb's constant in five dimensions to be completely free of ambiguities. 
Our results suggest that these expectations are indeed justified.

\section{$k_5$ from Gauge Theory}

Near the phase transition neither perturbation theory nor strong coupling techniques are useful. 
The appropriate analytical field theory computational tool instead is the Mean-Field expansion,
which can be conveniently implemented on a Euclidean lattice \cite{DZ}.
It is worth stressing that on the isotropic lattice both the order of the phase transition (it is of first order) and 
the numerical value of the critical coupling $\beta_c$ are predicted correctly by the Mean-Field \cite{DZ}, as
several Monte Carlo studies have proved \cite{Creutz}.
%
\begin{figure}[!t]
\centerline{\epsfig{file=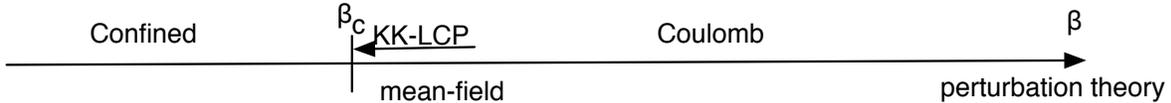,width=16cm}}
\caption{\small The phase diagram of isotropic, infinite volume five-dimensional $SU(N)$ gauge theories.}
\end{figure}
%

We begin by expressing eq. (\ref{5dC}) in lattice parameters as
\be
k^{(\rm LAT)}_5(L) = {\overline c}_2 \frac{\b}{N^2}\, ,\label{5dCL}
\ee
where ${\overline c}_2=c_2/a$ with $a$ the lattice spacing and $\b=2Na/g_5^2$ the dimensionless lattice coupling.
We take the number of lattice points $L=2\pi \rho/a$ in all five directions to be the same,
so $\rho$ is the physical radius of each periodic dimension.
On a finite lattice the quantity in eq. (\ref{5dCL}) depends on $L$.
To leading order in the Mean-Field expansion the computation of the Wilson Loop of length $r$ 
oriented along a four-dimensional slice, computed using Wilson's plaquette action, yields for $SU(2)$ the result \cite{MF}
\bea
aV_5(r/a)  &=& -2 \log{(\ov_0)}
-\frac{1}{\ov_0^2}\frac{1}{L^4}\sum_{p} \sum_{M\ne 0}
\d_{p_0,0}\, \Bigl\{ \bigl(\frac{1}{4}\cos(p_Mr)+1 \bigr)
K^{-1}_{00}(p,0) \nonumber\\
&+& \sum_A \bigl(\frac{1}{4}\cos(p_Mr)-1 \bigr)
K^{-1}_{00}(p,A)\Bigr\} \,.\label{StaticL}
\eea
In the above, $\ov_0$ is the Mean-Field background
(zero by definition in the confined phase, which explains why we did not consider a four-dimensional model) 
and $p=\{p_M=\frac{2\pi }{L}l_M\}, l_M=0,\cdots, L-1$ are the
lattice momenta. The Euclidean indices take the values $M,N=0,1,2,3,5$.
The inverse propagator $K_{MN}(p,\alpha)$, apart from $\b$ and $\ov_0$, depends on the momenta and the group index $\alpha=0,A=1,2,3$.
At this order, there is no distinction between bare and renormalized lattice coupling, so $\b$ 
can be thought of as a physical coupling.
Among the observables of this theory one finds states with vector and scalar quantum numbers.
Appropriate Polyakov Loops can be used to extract the lightest vector's mass, which turns out to be
$am_V=12.61/L$ in units of the lattice spacing and the lightest scalar's mass $am_S$ which depends only on $\b$ and therefore can be
used as a measure of the lattice spacing. All observables turn out to be gauge independent.
The related expressions with their detailed derivation can be found in \cite{MF}.
A Line of Constant Physics (LCP) is defined as the value
of a physical quantity along a trajectory of constant $q^{(LAT)}=m_V/m_S$. 
The LCP which corresponds to the scalar and vector having Kaluza-Klein masses is $q^{(LAT)}=1$.
We call this the KK-LCP.
The algorithm then is to compute eq. (\ref{StaticL}) numerically for a given $L$ and plot $(r/a)^2 aV_5(r/a)$ vs $(r/a)^2$.
This is expected to be a straight line whose intercept
is ${\overline c}_2$, 
from which our physical quantity $k^{(\rm LAT)}_5(L)$ can be easily obtained
\footnote{An alternative way to extract ${\bar c}_2$ would be from $1/2r^3 F_5(r)$, with $F_5(r)=\partial V_5(r)/\partial r$
by analogy to four dimensions \cite{FrancRainer}.
For our purposes here though a global fit will suffice.}. Then repeat these steps for several
increasing $L$'s along the KK-LCP (see Fig.1) and finally extrapolate to $k^{(\rm LAT)}_5(\infty)$. In Table 1 we present the available data.
For $SU(2)$ $\b_c\simeq 1.6762016760$ and from Table 1 we can see
how the lattice spacing decreases as the phase transition is approached.
\hskip 2cm
\begin{center}
\begin{tabular}{|c|c|c|c|}
\hline 
$L$ & $a m_S$ & $\b$ & $k^{(\rm LAT)}_5 \times 10^4$ \\
\hline \hline   
${\bf 24}$  & $0.5236$ &$1.6764598$ & $708$\\ \hline
${\bf 30}$  & $0.4189$ &$1.6763073$ & $700$\\ \hline
${\bf 36}$  & $0.3490$ &$1.67625254$ & $693$\\ \hline
${\bf 42}$  & $0.2992$ &$1.67622913$ & $687$\\ \hline
${\bf 48}$  & $0.2618$ &$1.67621776$ & $682$\\ \hline
${\bf 54}$  & $0.2327$ &$1.67621172$ & $678$\\ \hline
${\bf 60}$  & $0.2094$ &$1.67620825$ & $674$\\ \hline
${\bf 72}$  & $0.1745$ &$1.67620484$ & $668$\\ \hline
${\bf 96}$   & $0.1309$ &$1.676202674$ & $661$\\ \hline
${\bf 300}$ & $0.0419$ &$1.6762016769$ & $644$\\ \hline
\end{tabular}
\center{TABLE 1. The KK-LCP lattice data.}
\end{center}
The infinite volume extrapolation gives
\be
k_5^{(LAT)}(\infty)=0.0636\, ,
\ee
see Fig. 2. It is not easy to assign an error to our result because it is a numerical
computation of an analytical expression. A reasonable estimate gives an error of $\pm 0.0002$.

%
\begin{figure}[!t]
\centerline{\epsfig{file=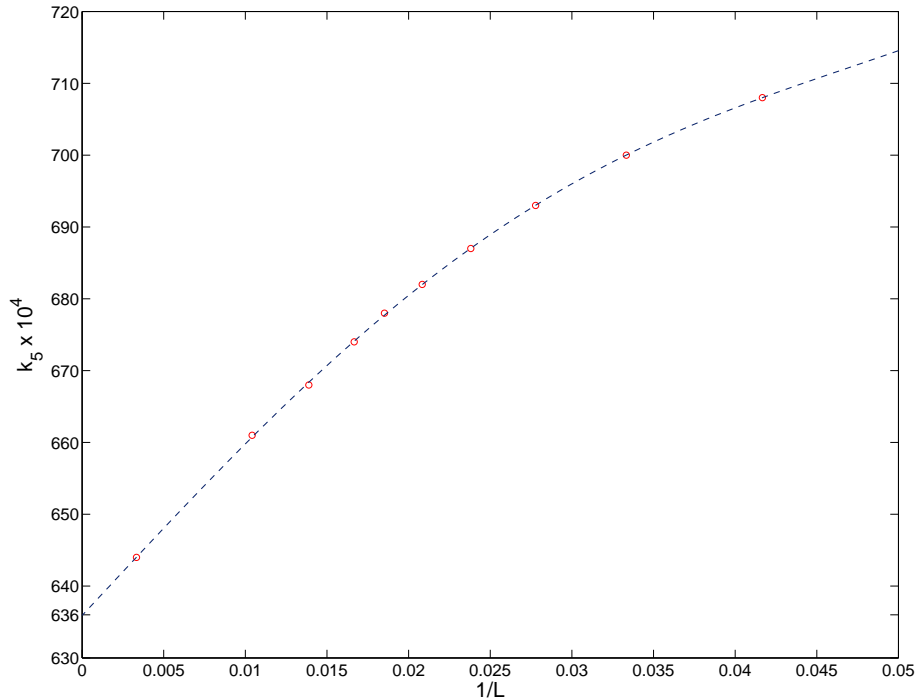,width=14cm}}
\caption{\small The infinite $L$ extrapolation of $k_5^{\rm (LAT)}(L)$. Since it is a first order phase transition, the physical volume goes
to infinity at a finite lattice spacing.}
\end{figure}
%

\section{$k_5$ from Gravity}

One of the reasons we chose to compute Coulomb's constant is because
we were after an $N$-independent quantity. 
Up to now we have computed it only for $N=2$ so we must prove that it is indeed $N$-independent.
Instead of developing the necessary Lattice Mean-Field formalism for general $N$ which 
is rather involved \cite{Kutsumb}, here we will try to prove it in a simpler way.

Consider a large number of coincident $N$ $D4$-branes in type IIA string theory with one of the spatial dimensions
along the brane compactified on a circle of radius $\rho$. The gravitational backreaction of such a configuration is \cite{Witten}
\bea
ds^2 = \left(\frac{u}{R}\right)^{3/2}
\left(\eta_{\mu\nu}dx^\mu dx^\nu+f(u)dx_5^{2}\right)
+\left(\frac{R}{u}\right)^{3/2}
\left(\frac{du^2}{f(u)}+u^2 d\Omega^2\right),
\eea
\be e^\phi= g_s \left(\frac{u}{R}\right)^{3/4},~F=\frac{2\pi N}{V}\epsilon
\ee
\be 
f(u)=1-\frac{u_k^3}{u^3} \ ,~R^3=\pi g_s l_s^3 N ~,
\ee
where $\mu,\nu=0,1,2,3$, $x_5$ parametrizes the circle and $d\Omega^2$, $\epsilon$ and $V=8\pi^2/3$
are the line element, the volume form and the volume of a unit $S^4$,
respectively. $R$ is the radius of $S^4$. $\phi$ is the dilaton, $F$ is the Ramond-Ramond 4-form
and $g_s$ and $l_s $ are the fundamental string coupling and length respectively.
$u_k$ is a minimal length scale in the $u$-direction, reflecting the absence of conformal invariance.
This background is believed to be dual to a five-dimensional gauge theory along $x_\mu, x_5$ compactified on
a circle. The low energy spectrum does not contain fermions 
because they are assumed to have anti-periodic boundary conditions along the circle,
but it does contain a set of adjoint scalars.
These however are expected to pick up nonperturbatively a large mass and eventually also decouple.
In this case, the dual theory at low enough energies is a pure, non-supersymmetric, five-dimensional $SU(N)$ gauge theory in the large
't Hooft coupling $\lambda$ limit, with $\lambda_5=4\pi^2 g_sl_s=2\pi \rho \lambda$.
The lightest massive state is the adjoint scalar that originates from the fifth-dimensional components of the
gauge field (not to be confused with the adjoint scalar mentioned above), with a Kaluza-Klein mass $1/\rho$.
It is known \cite{Maldacena} that the minimal surface of a string world-sheet 
parametrized by the coordinates $\sigma$ and $\tau$ extending in
the holographic $u$-dimension and whose boundary lies on the five-dimensional boundary of the 
six dimensional space transverse to the $S^4$,
represents the Wilson Loop of the dual gauge theory. 
More specifically, by taking the string world-sheet ansatz 
$t=\tau, x_1=\sigma, x_5=\pi \rho/2, u=u(\sigma)$,
one arrives at the expressions \cite{Sonni}
\be
r = 3\rho \sqrt{A}\int_1^\infty \frac{dy}{\sqrt{(y^3-A^3)(y^3-1)}}\label{L}
\ee
and
\bea
V_5 = \frac{u_k}{l_s^2} \frac{2}{A} \Biggl\{\int_1^\infty dy \Bigl[\frac{y^3}{\sqrt{(y^3-A^3)(y^3-1)}}
- \frac{1}{\sqrt{1-\frac{A^3}{y^3}}}\Bigr]-\int_A^1 dy \frac{1}{\sqrt{1-\frac{A^3}{y^3}}}\Biggr\}\;  \label{V5}
\eea
for the length of the Wilson Loop and the static potential respectively. 
The latter is computed in the regularization scheme where its finiteness is ensured by subtracting out
the infinite mass of the static quarks. 
We are using the dimensionless parameters $y=u/u_0$ and $A=u_k/u_0$
where $u_0\ge u_k$ is the turning point of the string world-sheet, i.e. the deepest the string probes the
holographic dimension. The $A\to 0$ limit corresponds to the ultra-violet limit of the gauge theory
where in fact the circle de-compactifies. 
In this limit the adjoint scalar of mass $1/\rho$ becomes part of the massless five-dimensional gauge field,
so $q^{(GRAV)}=m_V/m_S=1$ as in the previous section.
As $\rho$ increases and the system enters in its five-dimensional domain, $m_S$ goes faster
to zero compared to the mass of the other adjoint fields, as it is protected by gauge invariance.
Thus, as long as we are below its mass scale, the latter always remain decoupled. 
We can, instead of decompactifying the circle, stop at some large but finite $\rho$ where supersymmetry remains broken. 
In fact, large and  infinite $\rho$ are practically indistinguishable from the point of view of our observable.

In \cite{Giatagan} it was shown how to disentangle the system of eqs. (\ref{L}) and (\ref{V5}) in the ultraviolet, in order
to obtain $V_5(r)$. 
The solution involves series whose individual terms have divergences which however must cancel in the sum because
eq. (\ref{L}) is manifestly finite and eq. (\ref{V5}) is made finite by the regularization. 
In any case their effect accumulates in an irrelevant additive constant in $V_5$.
The final result is
\be
V_5(r) =  {\rm const.} -\frac{c_2}{r^2}, \;\; c_2 = \frac{1}{54\pi^2}\left(\frac{\sqrt{\pi}\G(2/3)}{\G(7/6)}\right)^3 \lambda_5
\, .\label{d5}
\ee
It is possible to verify this numerically, directly from eqs. (\ref{L}) and (\ref{V5}), 
(that also shows that the additive constant actually sums to zero).
This is a five-dimensional Coulomb potential that points to the dual gauge theory being in 
its Coulomb phase. Moreover, the duality works at strong coupling
so the gauge theory must be strongly coupled.
These requirements are simultaneously satisfied near the phase transition, precisely in the regime where we
computed the static potential on the lattice.
In order to make quantitative contact with the lattice formulation, we rewrite eq. (\ref{d5})
in terms of lattice parameters as
\be
{\overline c}_2 = \frac{1}{27\pi^2}\left(\frac{\sqrt{\pi}\G(2/3)}{\G(7/6)}\right)^3\frac{N^2}{\b}\, .\label{c2dimless}
\ee
Eq. (\ref{c2dimless}) implies that the obvious $N$-independent dimensionless quantity
that plays the role of Coulomb's constant is
\be
k_5^{(GRAV)} = {\overline c}_2\frac{\beta}{N^2}\label{k5G}
\ee
in agreement with eq. (\ref{5dCL}) and suggests for it the value
\be
k_5^{(GRAV)} =\left[\frac{B(2/3,1/2)}{3\pi^{2/3}}\right]^3=0.0649\cdots
\ee
with $B(x,y)$ the Euler Beta function.
The discrepancy between $k_5^{(LAT)}$ and $k_5^{(GRAV)}$ is $1.81\%-2.43\%$.

\section{Conclusion}

Motivated by its four-dimensional analogue, we proposed a definition for Coulomb's constant 
for five-dimensional $SU(N)$ gauge theories and then computed it in two different ways.
One in a Lattice Mean-Field expansion for $N=2$ and the other
via a Holographic Wilson Loop calculation.
The agreement of the results, numerically found to be within approximately $2\%$,
relies on the validity of the duality between the two approaches. 
Even though the duality is expected to hold for large $N$,
we argued that the comparison makes sense because  
-as the numerical agreement suggests- we are dealing with 
a quantity that seems to remain $N$-independent to a good approximation beyond the large $N$ limit.

\vskip .1cm
{\bf Acknowledgements}

The author is grateful to F. Knechtli for many useful comments on a draft and 
for generating the data for the larger lattices at the U. Wuppertal.
He would also like to thank  the KITP, Santa Barbara and especially the organizers of the
program "Novel Numerical Methods for Strongly Coupled Quantum Field Theory and Quantum Gravity" for hospitality.
This research was supported in part by 
the NTUA grants PEBE 2009, 2010 and in part by the National Science Foundation under Grant No. PHY11-25915.
 \newpage


\end{document}